\documentclass[twocolumn]{aastex631}
\usepackage{lineno}

\DeclareOldFontCommand{\bf}{\normalfont\bfseries}{\mathbf} 
\providecommand{\DIFdel}[1]{} 
\providecommand{\DIFdelbegin}{} 
\providecommand{\DIFdelend}{} 
\providecommand{\DIFdelFL}[1]{\DIFdel{#1}} 
\providecommand{\DIFdelbeginFL}{} 
\providecommand{\DIFdelendFL}{} 
\RequirePackage{listings} 
\lstdefinelanguage{DIFcode}{ 
  moredelim=[il][\color{white}\tiny]{\%DIF\ <\ }, 
  moredelim=[il][\sffamily\bfseries]{\%DIF\ >\ } 
} 
\lstdefinestyle{DIFverbatimstyle}{ 
	language=DIFcode, 
	basicstyle=\ttfamily, 
	columns=fullflexible, 
	keepspaces=true 
} 
\lstnewenvironment{DIFverbatim}{\lstset{style=DIFverbatimstyle}}{} 
\lstnewenvironment{DIFverbatim*}{\lstset{style=DIFverbatimstyle,showspaces=true}}{} 

\begin{document}
\title{Nightside Clouds on Tidally-locked Terrestrial Planets Mimic Atmosphere-Free Scenarios}

\author[0000-0002-4250-0957]{Diana Powell}
\affiliation{Department of Astronomy \& Astrophysics, University of Chicago, Chicago, IL 60637, USA}

\author{Robin Wordsworth}
\affiliation{School of Engineering and Applied Sciences, Harvard University, Cambridge, MA, USA}
\affiliation{Department of Earth and Planetary Sciences, Harvard University, Cambridge, MA, USA}

\author{Karin Öberg}
\affiliation{Center for Astrophysics Harvard \& Smithsonian,
Cambridge, MA 02138, USA}

\begin{abstract}
We investigate the impact of nightside cloud formation on the observable night-day contrast of tidally-locked terrestrial planet atmospheres. We demonstrate that, in the case where the planetary dayside is only 10s of Kelvin hotter than the planetary nightside, the presence of optically thick nightside clouds can lead to observations that mimic a planet without an atmosphere, despite the planet actually hosting a significant (10 bar) atmosphere. The scenario presented in this work requires a level of intrinsic atmospheric day/night temperature contrast such that the nightside can form clouds while the dayside is too hot for cloud formation to occur. This scenario is most likely for hotter terrestrials and terrestrials with low volatile inventories. We thus note that a substantial dayside/nightside temperature difference alone does not robustly indicate that a planet does not host an atmosphere and additional observations and modeling are essential for characterization. We further discuss several avenues for future study to improve our understanding of the terrestrial planets and how best to characterize them with JWST.  
\end{abstract}


\section{Introduction}
The majority of known terrestrial planets orbit M dwarf stars \citep{Dressing2015,2017AJ....154..109F}. Due to detection limits, most known terrestrial planets orbit close to their host stars such that they are likely tidally locked with permanent daysides and permanent nightsides. The increased longevity of stellar activity of the host M dwarf stars, which may remain active for $\sim$2-3 Byr \citep{Medina2022}, may make these planets more susceptible to atmospheric loss \citep[e.g.,][]{do2022}. Thus, it is unclear whether terrestrial planets in close-in orbits around M stars can maintain or replenish their atmospheres over geologic time. 

\DIFdelbegin \DIFdel{To date, five terrestrial planets , LHS 3844b, GJ 1252b, Trappist 1b, Trappist 1c and GJ 367b, }\DIFdelend An increasing number of terrestrial planets have been detected in emission (either secondary eclipse or phase curves)\citep[e.g.,][]{kreidberg2019,crossfield2022,Greene2023,zieba2023,zhang2024} . These observations have indicated the existence of strong day/night temperature gradients across the planet such that these planets are thought to have hot daysides and cool nightsides. Simplified atmospheric models indicate that these observations are indicative of planets that have little or no planetary atmosphere able to redistribute radiation from the host star from the dayside to the nightside of the planet. Thus, taken together, these results point to the efficient removal of close-in terrestrial planet atmospheres. 

However, \DIFdelbegin \DIFdel{the models presented in previous works have neglected the }\DIFdelend previous models that include radiative effects of spatially inhomogeneous atmospheric clouds \DIFdelbegin \DIFdel{. While some models have considered clouds in a global sense, the spatial variations in cloudiness across the planet can be extremely important in shaping atmospheric observations}\DIFdelend have primarily focused on temperate terrestrials with substantial volatile inventories \citep[e.g.,][]{yang2013, Haqq-Misra2018,Komacek2019} leaving much of the potential parameter space of known terrestrial planets unexplored. The impact of inhomogeneous aerosols on atmospheric spectra has already been shown to be empirically important in the substantially different case of hot Jupiters, which are tidally locked giant planets with extreme insolation gradients and day/night temperature contrasts. In particular, hot Jupiters have constant nightside emission temperatures across a range of planetary equilibrium temperatures while the same is not true for their dayside emission temperature which increases monotonically with planetary equilibrium temperature when observed at 3.6 and 4.5 microns \citep{beatty2019,keating2019}. This effect has been shown to be a consequence of abundant nightside clouds \citep{Gao2021,parmentier2021}. In essence, the nightside observations always probe the emission temperature at the top of the optically thick nightside cloud layer, which is located at the same temperature across a range of planetary equilibrium temperatures even though the pressure-level of the cloud top decreases by more than 3 orders of magnitude (from $\sim$1 to $\sim$10$^{-3}$ bar) with increasing planetary equilibrium temperatures \citep{Gao2021}. On the hotter planetary daysides, clouds either do not form in the photosphere or they do not form as abundantly such that the dayside emission temperature is set by the gas opacity (primarily the water opacity) which reaches unity at a hotter temperature for planets with higher equilibrium temperatures. The net result is a substantially larger observed dayside-nightside temperature contrast than the actual temperature contrast. In addition to producing an exacerbated temperature contrast, the nightside cloud layer can also increase the actual dayside temperature, since nightside clouds on \DIFdelbegin \DIFdel{ho }\DIFdelend hot Jupiters blanket the nightside atmosphere thus blocking nightside re-radiation and correspondingly increasing the temperature of the planetary dayside \citep{parmentier2021}. Thus, the observed day/night brightness temperature contrasts in hot Jupiters is exaggerated compared to the true day/night temperature contrast and this is particularly true for the most highly irradiated planets. 

In this work we show that a similar effect in the atmospheres of \DIFdelbegin \DIFdel{highly irradiated }\DIFdelend tidally-locked terrestrial planets could cause a planet with a substantial atmosphere to appear as though there is no significant redistribution of radiation from the planetary dayside to the planetary nightside, thus appearing as though it lacks an atmosphere. In particular, we show that even a modest day/night temperature difference can lead to a substantial exaggeration in observed day/night temperature contrasts. In Section \ref{day_night}, we discuss reasonable day/night temperature contrasts on highly irradiated terrestrial planets that host substantial atmospheres. In Section \ref{night_clouds}, we discuss the planetary conditions where this effect is most likely to be relevant and describe the impact of nightside clouds on the observed nightside emission temperature. In Section \ref{clouds_daynight}, we describe the impact of a cloud-free planetary dayside coupled with a cloudy planetary nightside on the observed day/night temperature contrast. In Section \ref{diverse_clouds}, we briefly discuss the potential phase space of clouds on terrestrial planets. We discuss our findings and avenues for future study in Section \ref{discuss} and conclude in Section \ref{conclude}.  

\begin{figure} 
   \centering
   \includegraphics[width=0.39\textwidth]{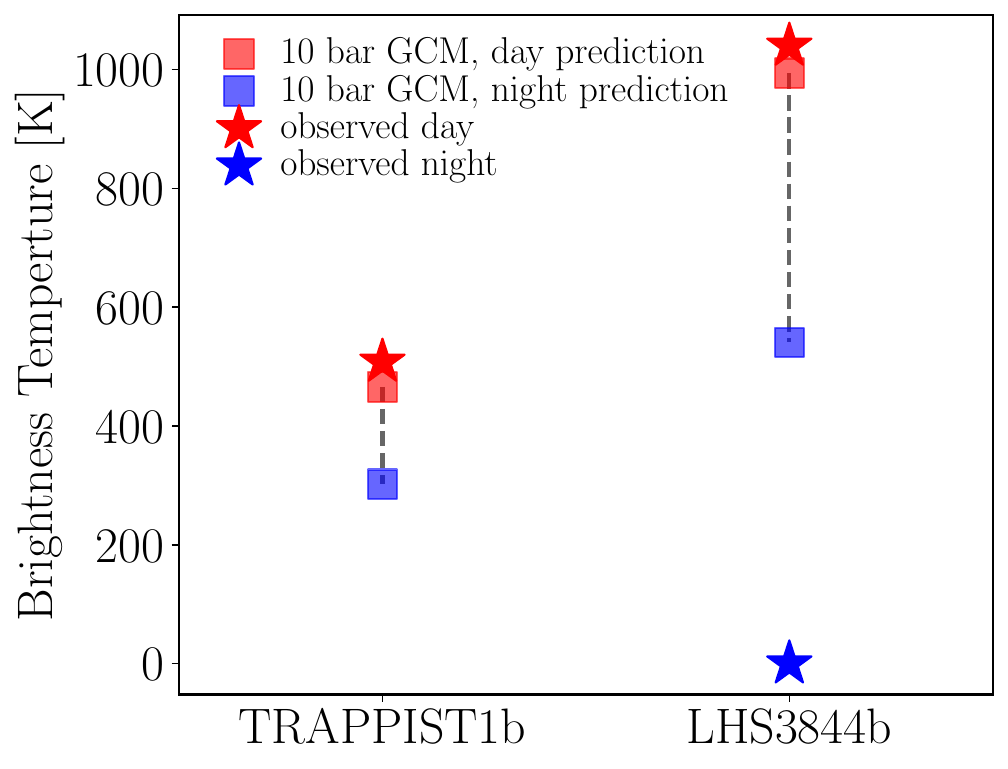}
   \caption{Even in the most optimistic case of an optically thin atmosphere ($\tau = 0.1$), GCM predictions from \citet{koll2022} for dayside and nightside brightness temperatures for Trappist 1b and LHS 3844b are inconsistent with observations. This is particularly true in regards to LHS 3844b where no nightside emission was detected from the planet.}
   \label{fig:predict_observed}
\end{figure}

\section{Day-Night Temperature Contrasts on Terrestrial Planets}\label{day_night}
The day/night temperature contrast for highly-irradiated tidally-locked terrestrial planets with substantial atmospheres is an active area of research as the phase space of potential atmospheric compositions and properties is vast. However, previous models of tidally locked terrestrial planets give insight into the approximate temperature difference between the dayside and nightside. 

In the cloud-free GCM simulations presented in \citep{koll2022}, the observed difference in dayside/nightside brightness temperatures for a planet with a 10 bar atmosphere can range from roughly 500 K to 100 K, depending on the optical depth at the surface of the planet, for LHS 3844b (period = 0.46 days). In the same work the difference in dayside/nightside brightness temperature for TRAPPIST-1b (period = 1.51 days) can range from roughly 200 K to a few K again depending on the optical depth at the surface of the planet. Simulations of temperate tidally-locked planets in the habitable zones of M-dwarf stars have been shown to have day/night temperature contrasts of $\sim$10-60 K (here referring to atmospheric temperature, not brightness temperature) particularly at the planetary surface \citep{Haqq-Misra2018, Komacek2019,joshi-etal-1997,Wordsworth2015}. Even terrestrial bodies that are not tidally locked, such as the Earth, can have $\sim$ 20 K differences in temperature between the day and night sides \citep{pierrehumbert-2010, Komacek2019}. 

These studies thus indicate that there can be substantial differences in the actual day/night temperature contrast as well as the observed brightness temperature contrast for terrestrial planets with substantial atmospheres. However, these differences in temperature are small compared to the large brightness temperature contrasts observed for tidally locked terrestrial planets. For example, in the case with a 10 bar atmosphere that is optically thin, the nightside temperature is predicted to be significantly higher than observed for LHS 3844b \citep{kreidberg2019}. The implication is that \DIFdelbegin \DIFdel{the planets lack }\DIFdelend this planet lacks an atmosphere. Two examples of this incongruity are shown in Figure \ref{fig:predict_observed}. As shown, while the dayside brightness temperatures are roughly consistent, even in the most optimistic case where a 10 bar atmosphere is optically thin, the nightside brightness temperature of LHS 3844b should be significantly higher than observed. 

\DIFdelbegin 
{
\DIFdelFL{Even in the most optimistic case of an optically thin atmosphere ($\tau = 0.1$), GCM predictions from \citet{koll2022} for dayside and nightside brightness temperatures for Trappist 1b and LHS 3844b are inconsistent with observations. This is particularly true in regards to LHS 3844b where no nightside emission was detected from the planet.}}
\DIFdelend \section{The Case of Nightside-Only Clouds on Tidally Locked Terrestrials}\label{night_clouds}

\DIFdelbegin 
{
\DIFdelFL{Spatially inhomogeneous nightside clouds can mimic atmosphere-free terrestrial planets. (a) A cartoon picture of a terrestrial planet without an atmosphere. The observed brightness temperature comes from thermal flux from the hot dayside surface and cold nightside surface. This may appear similar to (b) the case where the planet hosts a substantial optically thin atmosphere and an optically thick layer of nightside clouds. Both the day and nightside of the planet are similarly hot and the observed brightness temperature again originates near the surface of the dayside, however, the emission from the nightside arises from above the optically thick cloud deck. In this picture, the observed day/night brightness temperature contrast can be much larger than the intrisic day/night temperature contrast. }}

\DIFdelend We now consider under which conditions a cloudy night side and clear day side may occur on a tidally-locked terrestrial planet. Based on general circulation models of tidally-locked terrestrial planets, the dominant atmospheric flow pattern for slow rotators is predicted to be a large-scale day to night side wind (with a corresponding night to day side flow at lower altitudes) \citep[see e.g., ][]{Wordsworth2022}. In this case, hot atmospheric gases from the dayside cool as they are transported to the nightside. 
\DIFdelbegin \DIFdel{In the case where the atmosphere hosts a condensible species, there is a potential region of planetary thermodynamic }\DIFdelend 

In this work, we model a particular region of phase space where the atmosphere  is too hot on the dayside for a cloud species to condense but becomes cool enough for abundant cloud formation as material is transported to the planetary nightside. This is in contrast to the classic picture of volatile-rich, temperate terrestrials where moist gas is supersaturated on the planetary dayside and leads to abundant dayside clouds \citep[e.g.,][]{pierrehumbert2019,yang2014water} \citep[we note that simulations of drier planets have been performed though the focus was not on cloud formation, e.g.,][]{leconte-etal-2013,Hammond2017,ding2020}. We focus on planets with daysides where the condensible species is undersaturated such that the dayside remains clear. This region of phase space is most likely to occur for hotter terrestrials and/or terrestrials with depleted volatile inventories.

\begin{figure} 
   \centering
   \includegraphics[width=0.39\textwidth]{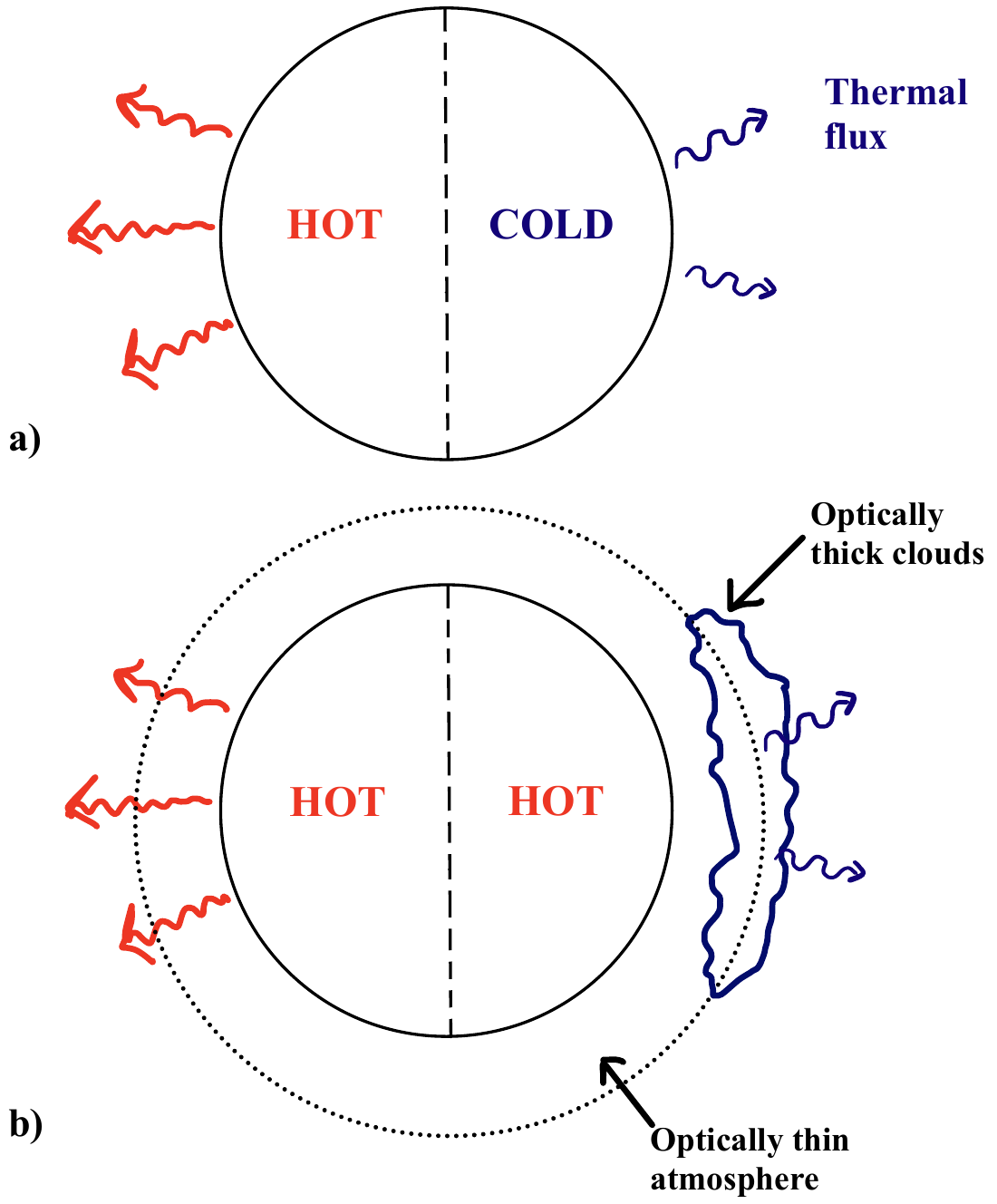}
   \caption{Spatially inhomogeneous nightside clouds can mimic atmosphere-free terrestrial planets. (a) A cartoon picture of a terrestrial planet without an atmosphere. The observed brightness temperature comes from thermal flux from the hot dayside surface and cold nightside surface. This may appear similar to (b) the case where the planet hosts a substantial optically thin atmosphere and an optically thick layer of nightside clouds. Both the day and nightside of the planet are similarly hot and the observed brightness temperature again originates near the surface of the dayside, however, the emission from the nightside arises from above the optically thick cloud deck. In this picture, the observed day/night brightness temperature contrast can be much larger than the intrinsic day/night temperature contrast. }
   \label{fig:cartoon}
\end{figure}

In this regime, there is a case where the atmospheric dayside is cloud-free while the nightside boasts a substantial, potentially optically-thick, cloud layer. This is illustrated in Figure \ref{fig:cartoon}. This picture is analogous to inhomogeneous cloud cover on hot Jupiters which primarily arises due to local differences in atmospheric thermal structure \citep[e.g.,][]{parmentier2016transitions,2018ApJ...860...18P,2019ApJ...887..170P,Gao2021,parmentier2021}. \DIFdelbegin \DIFdel{Thus, this picture of cloud formation differs from the case where a condensible species is supersaturated across the entirety of the planetary atmosphere, as is the case for many terrestrial exoplanet modeling studies where clouds form first on the dayside via moist adiabatic cooling \citep[e.g.,][]{pierrehumbert2019}}\DIFdelend Similarly, for faster rotators, the large-scale dynamics may support a strong equatorial jet which could similarly lead to cloud formation patterns reminiscent of those on hot Jupiters.  The picture of cloud formation discussed in this work relies on some level of temperature difference between the atmospheric dayside and nightside that allows for a marginally supersaturated \DIFdelbegin \DIFdel{condensible }\DIFdelend condensable species in the upper atmosphere of the planetary nightside while that same \DIFdelbegin \DIFdel{condensible }\DIFdelend condensable species is undersaturated on the planetary dayside \citep[as is shown for GJ1132b in][]{ding2019}. Thus, both the day/night temperature contrast and the total atmospheric volatile inventory \DIFdelbegin \DIFdel{\citep[that sets the level of condensible saturation, e.g.,][]{ding2022} }\DIFdelend are the key controlling parameters of cloud formation in these atmospheres.

 We also note that it may be possible to have clouds present on only the atmospheric nightside without any intrinsic difference in planetary dayside/nightside temperature. This has been modeled for the case of early Venus (see \citet{Turbet2021}) where clouds are primarily located on the nightside of the planet due to atmospheric dynamics alone. Thus, the picture presented in this work may be more generally applicable and future study that incorporates atmospheric dynamics, cloud microphysics, and self-consistent radiative transfer across a broad range of planetary parameters will better clarify the likelihood of the proposed scenario.  

\begin{figure} 
   \centering
   \includegraphics[width=0.5\textwidth]{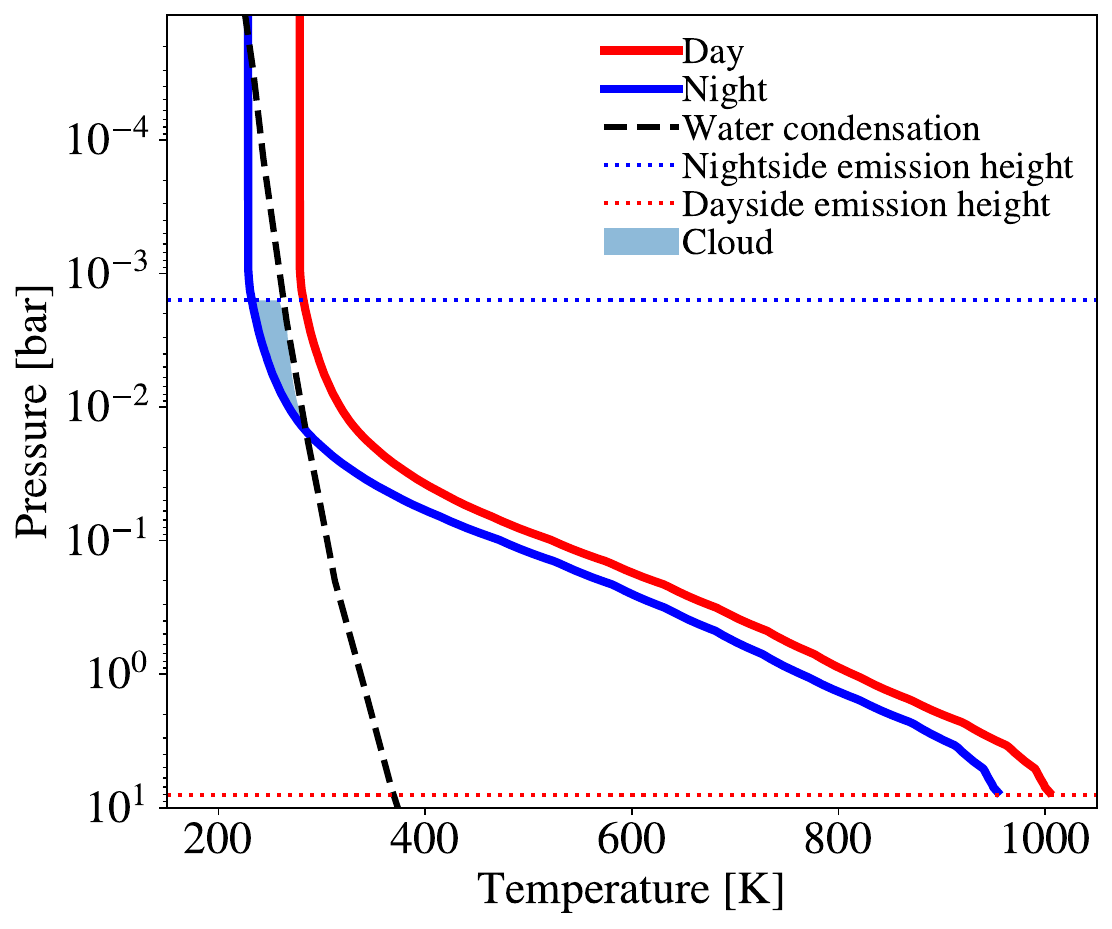}
   \caption{Clouds alter the location where the atmosphere emits. In this case \DIFdelbeginFL \DIFdelFL{we }\DIFdelendFL the dayside atmospheric temperature/pressure profile of an idealized terrestrial planet is only 50 K hotter (at all pressures) than the nightside of the same planet. However, the nightside of the planet is cool enough to form clouds in the upper atmosphere. The dayside emission height will depend on the atmospheric composition and resulting gas-phase opacity. In the most extreme case where the atmospheric gases are optically thin, the dayside emission height will be at the planetary surface. However, the nightside emission height will be at the top of the cloud deck if the clouds are optically thick. }
   \label{fig:temp}
\end{figure}

Clouds on the nightside of hot terrestrial planets have the potential to significantly alter the observed nightside emission spectra of a planet. In the case where the nightside cloud layer is optically thick, the thermal emission from the nightside of the planet will arise from the cloud top. This is demonstrated for a fiducial case in Figure \ref{fig:temp} where a small temperature difference between the day and night side results in a night side temperature that drops below the water condensation temperature, while the day side temperature does not. In our fiducial case we consider a 10 bar atmosphere and impose a 50K difference in day/night temperature structure due to inefficiencies in atmospheric heat redistribution alone, which we consider to be a conservative case given the discussion in Section \ref{day_night}. We note that the day/night temperature difference is unlikely to be constant with pressure and is instead more likely to have a smaller difference deeper in the atmosphere until very near the surface where an inversion may develop \citep[e.g.,][]{Wordsworth2015}. For this work, we are primarily interested in day/night temperature differences in the upper atmosphere which produce the largest observed day/night temperature contrast. The thermal structure of the atmosphere follows a dry adiabat with an isothermal region in the upper atmosphere which is informed from studies of tidally-locked hot Jupiters \citep[e.g.,][]{parmentier20133d} and hotter terrestrials \DIFdelbegin \DIFdel{\citep[e.g.,][]{malik2019}}\DIFdelend \citep[e.g.,][]{Wordsworth2015,malik2019} (which tend to have either an inversion or isothermal region in the upper atmosphere). Thus, the case of nightside-only cloud formation may occur even with modest differences in day/night temperature contrasts, on the order of or less than those discussed in Section \ref{day_night}.

Assuming that the kinetic conditions of cloud formation are favorable, an optically thick cloud layer could form in the upper regions of the atmosphere on the planetary nightside. In this case, the nightside emission height would be close to $\sim10^{-3}$ bar where the atmosphere approaches $\sim$ 200 K in temperature. The atmospheric nightside brightness temperature would therefore originate from the top of the optically thick cloud deck and would therefore would be very low. 

\begin{figure} 
   \centering
   \includegraphics[width=0.49\textwidth]{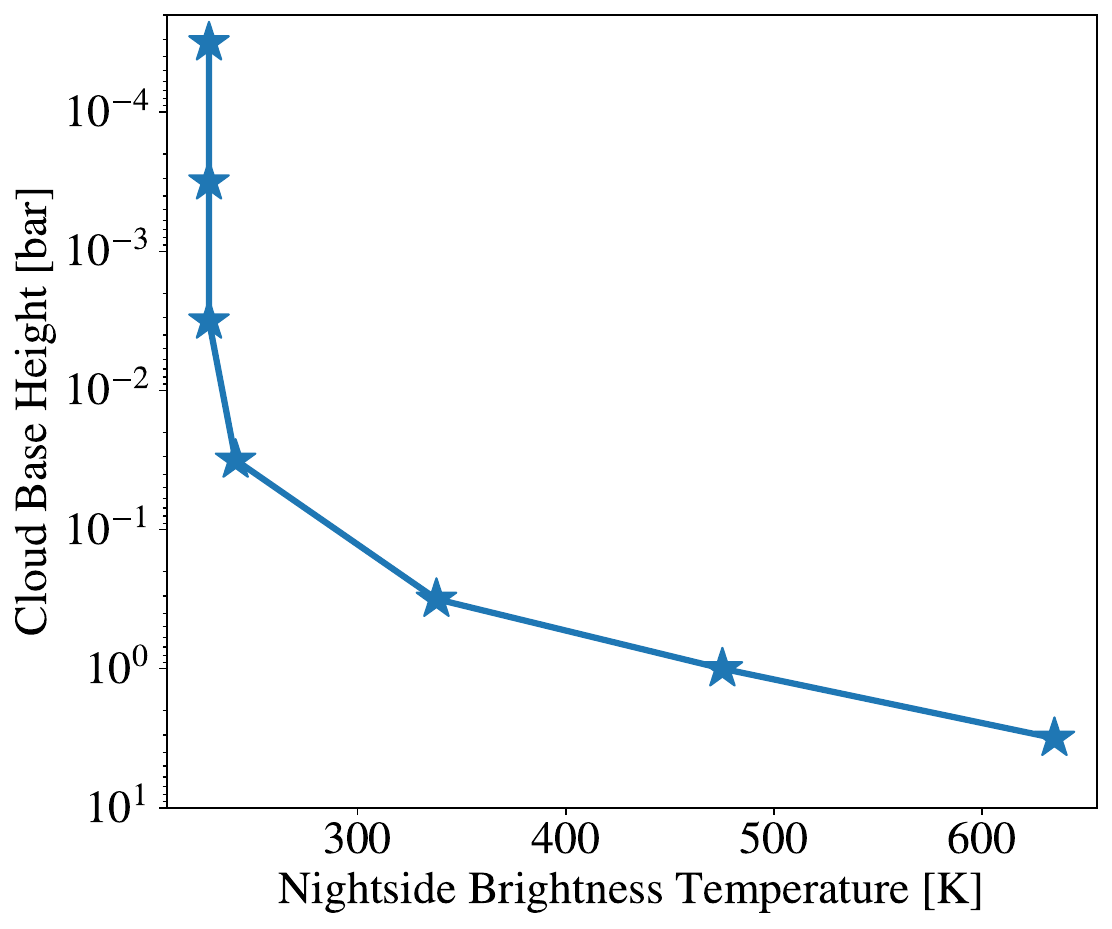}
   \caption{The height of the cloud deck alters the nightside emission height and thus the observed nightside brightness temperature. Here we vary the location of a gray cloud deck that extends over 1 \DIFdelbeginFL \DIFdelFL{bar }\DIFdelendFL dex of atmospheric pressure. Clouds present in the upper atmosphere give rise to low observed brightness temperatures. }
   \label{fig:nightside}
\end{figure}

 More generally, in the case of an optically thick nightside cloud deck the observed nightside brightness temperature will vary as a function of the height of the cloud base and how this relates to the atmospheric thermal structure. This can be seen in Figure \ref{fig:nightside} for the fiducial atmosphere model in Figure \ref{fig:temp} where we have imposed an optically thick cloud deck that extends one dex in pressure and have simply lowered the base of the cloud deck in the atmosphere. In this case, clouds in the upper atmosphere lead to low nightside brightness temperatures while clouds in the lower atmosphere lead to high nightside brightness temperatures. This can be understood because clouds present in the deeper atmosphere (with a fixed cloud deck height) will push the emission surface to higher pressures and temperatures. Therefore, for atmospheres that are cooler at higher altitudes, clouds present in the upper atmosphere derive the coolest brightness temperatures.  

\begin{figure*} 
   \centering
   \includegraphics[width=0.49\textwidth]{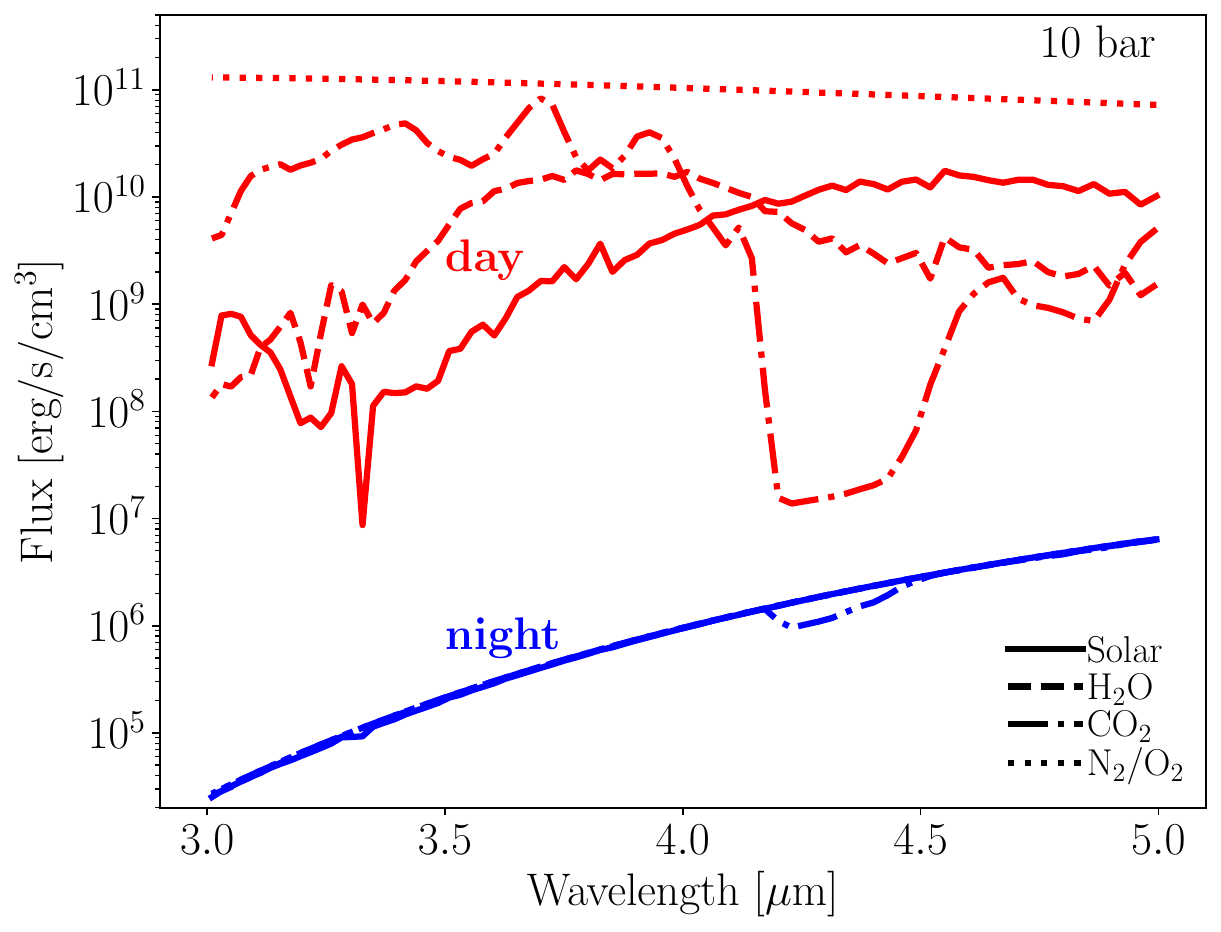}
   \includegraphics[width=0.49\textwidth]{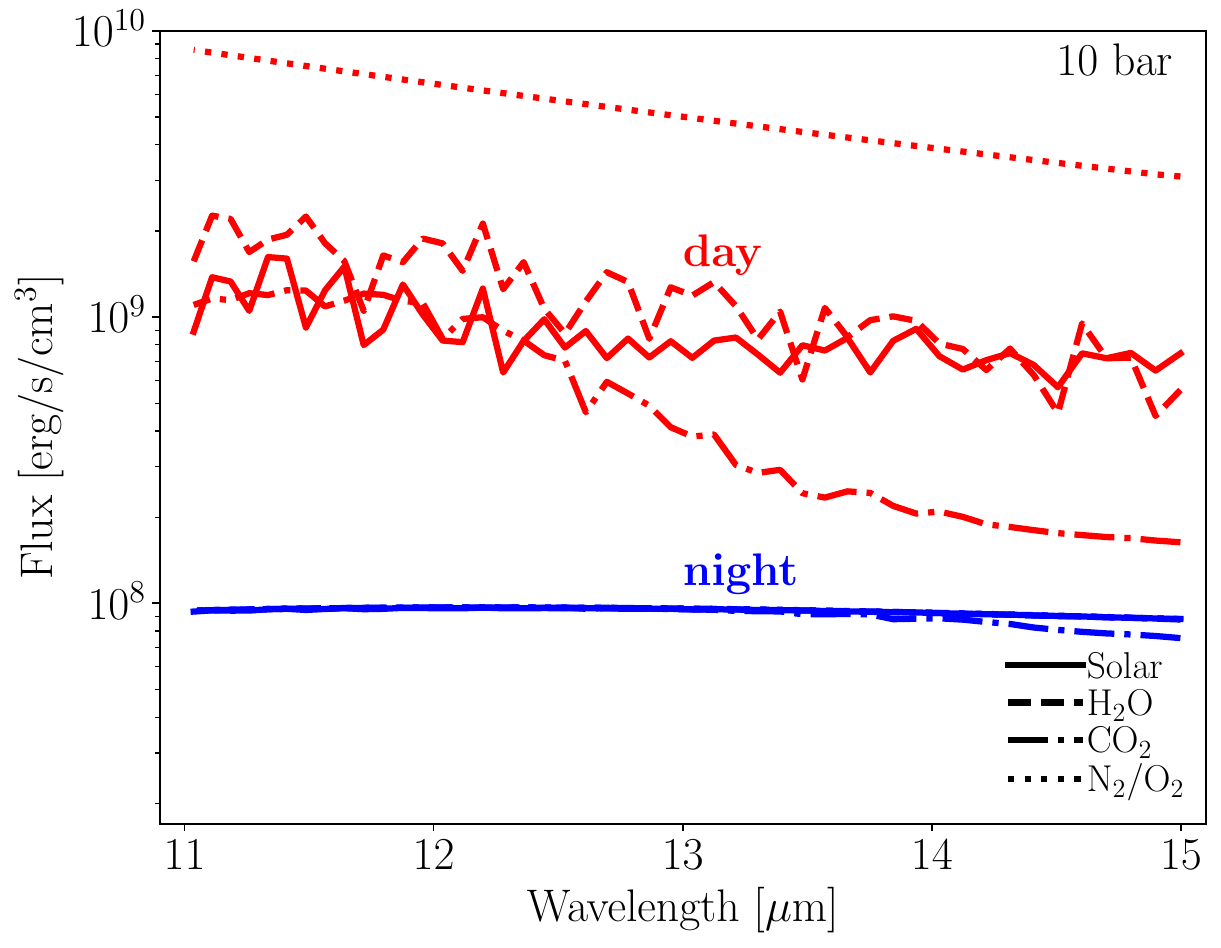}
   \includegraphics[width=0.49\textwidth]{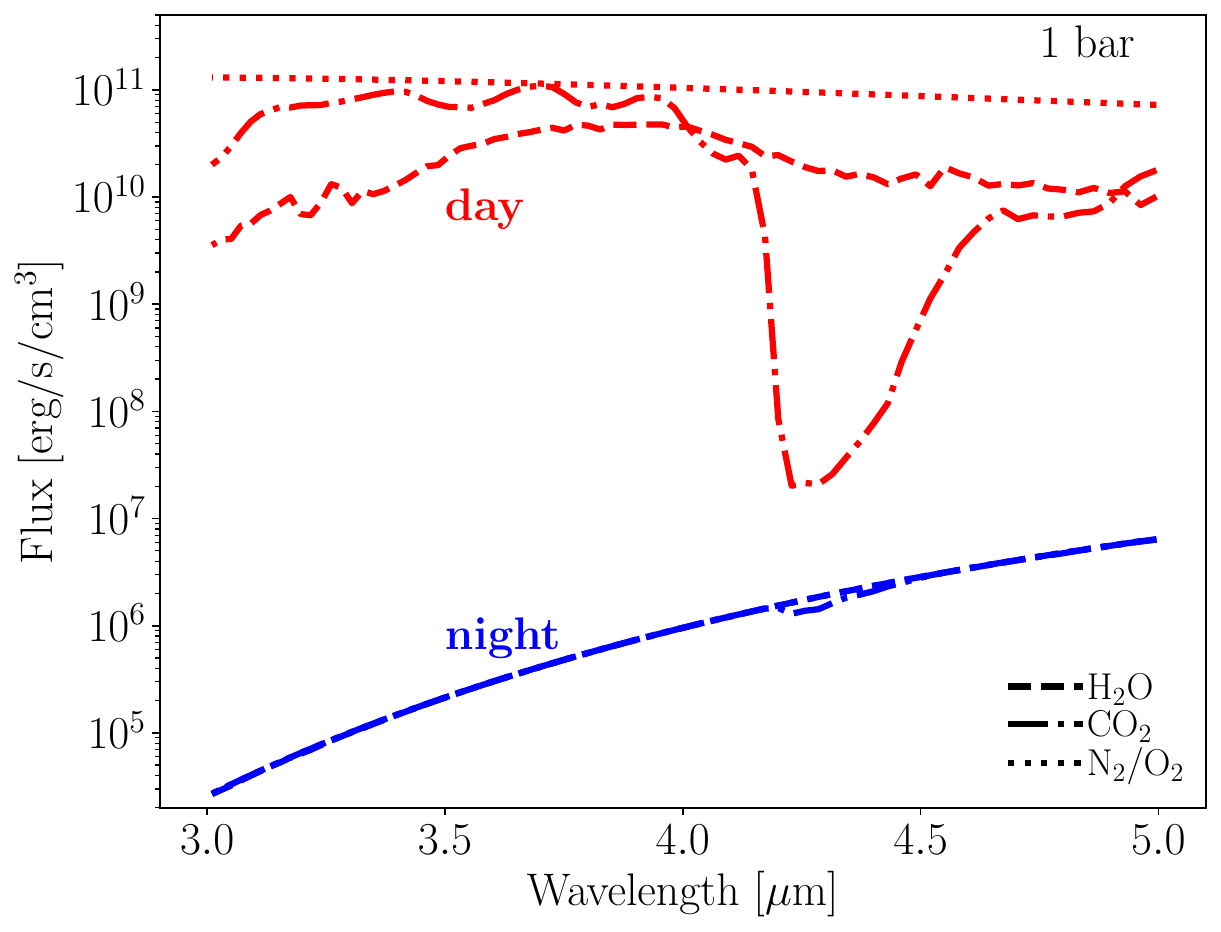}
   \includegraphics[width=0.49\textwidth]{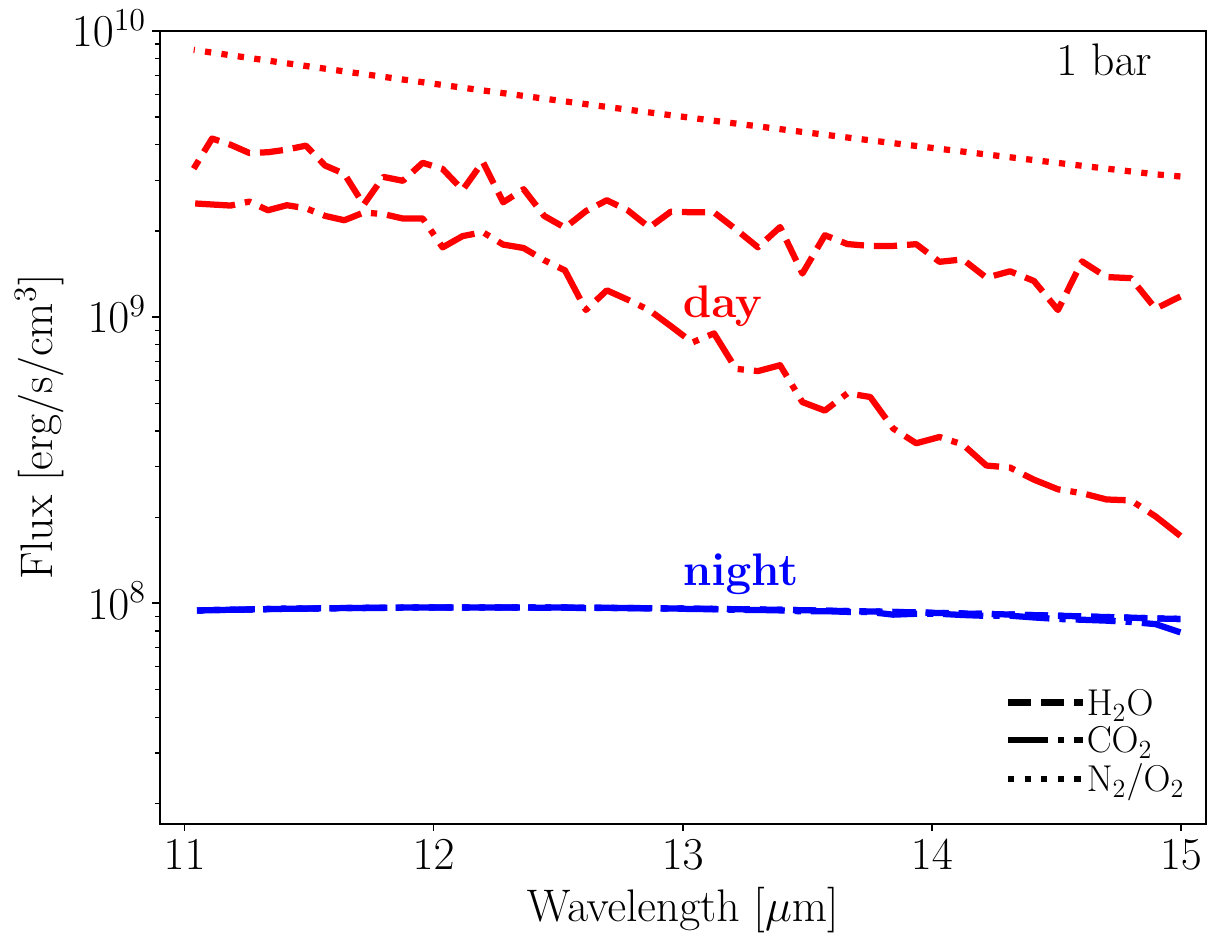}
   \caption{The atmospheric flux from the planetary dayside (red) and nightside (blue) for a host of atmospheric compositions for the case of a 10 bar atmosphere (top) and 1 bar atmosphere (bottom). In most cases (except for an N$_2$ and O$_2$ dominated atmosphere) the dayside emission exhibits atmospheric features while the nightside emission resembles a blackbody emitting at a low temeprature). In all cases nightside clouds give rise to significant dayside/nightside temperature contrasts.}
   \label{fig:flux}
\end{figure*}

Conversely, in the case where the atmospheric dayside is clear and the atmosphere is optically thin, the emission surface may be close to the surface of the planet where the atmospheric temperature is substantially hotter (for more optically-thick atmospheres the emission surface would be higher in the atmosphere). In such a case, the dayside brightness temperature could be up to the surface temperature of the planet, set by the incident stellar irradiation and the albedo of the planetary surface. In the fiducial case in Figure \ref{fig:temp}, the dayside brightness temperature could approach $\sim$1000 K. 

Thus, if the dayside of a terrestrial planet is clear while the nightside hosts an optically thick cloud layer, the emission will arise from very different regions of the atmosphere with correspondingly different thermal structures. As such, clouds on the planetary nightside have the potential to dramatically exacerbate observed day/night temperature contrasts on tidally locked terrestrial planets. 

\begin{figure*} 
   \centering
   \includegraphics[width=0.49\textwidth]{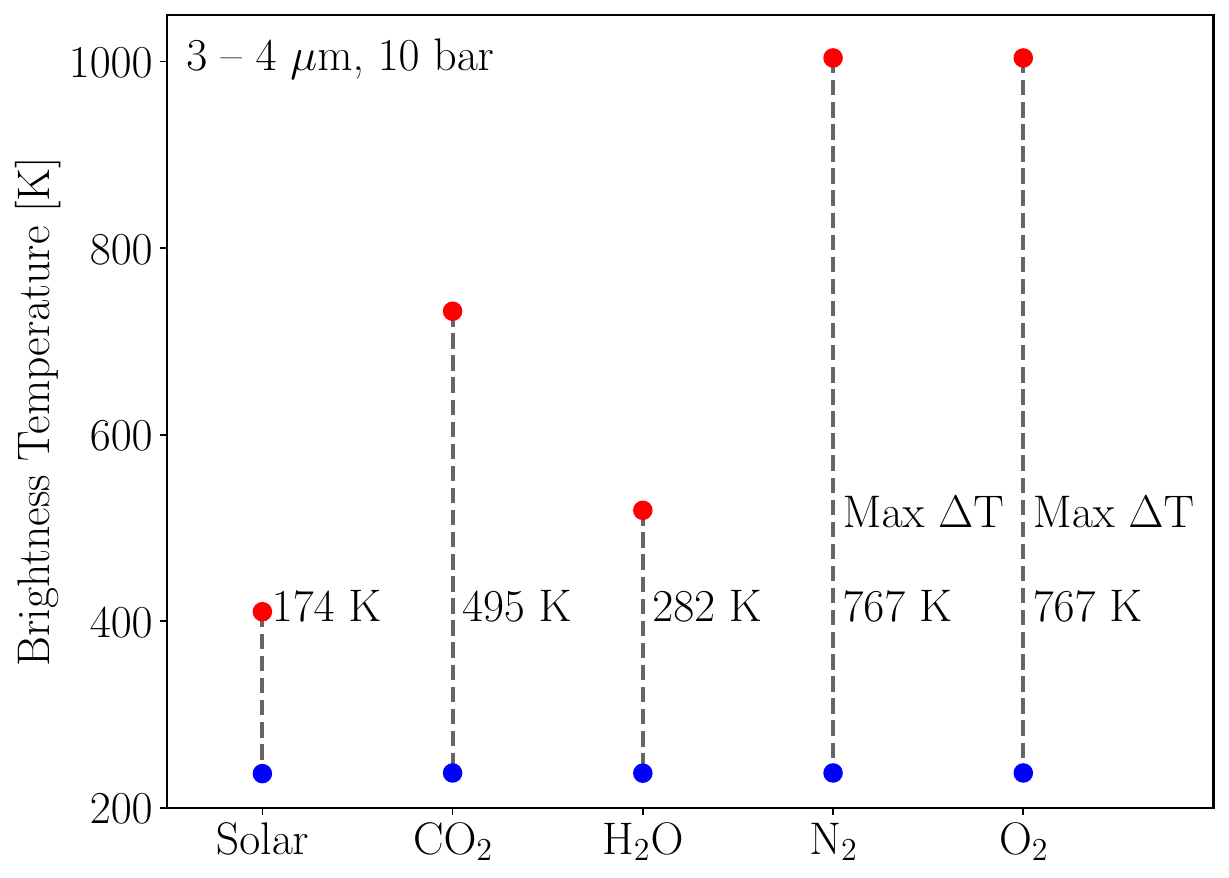}
   \includegraphics[width=0.49\textwidth]{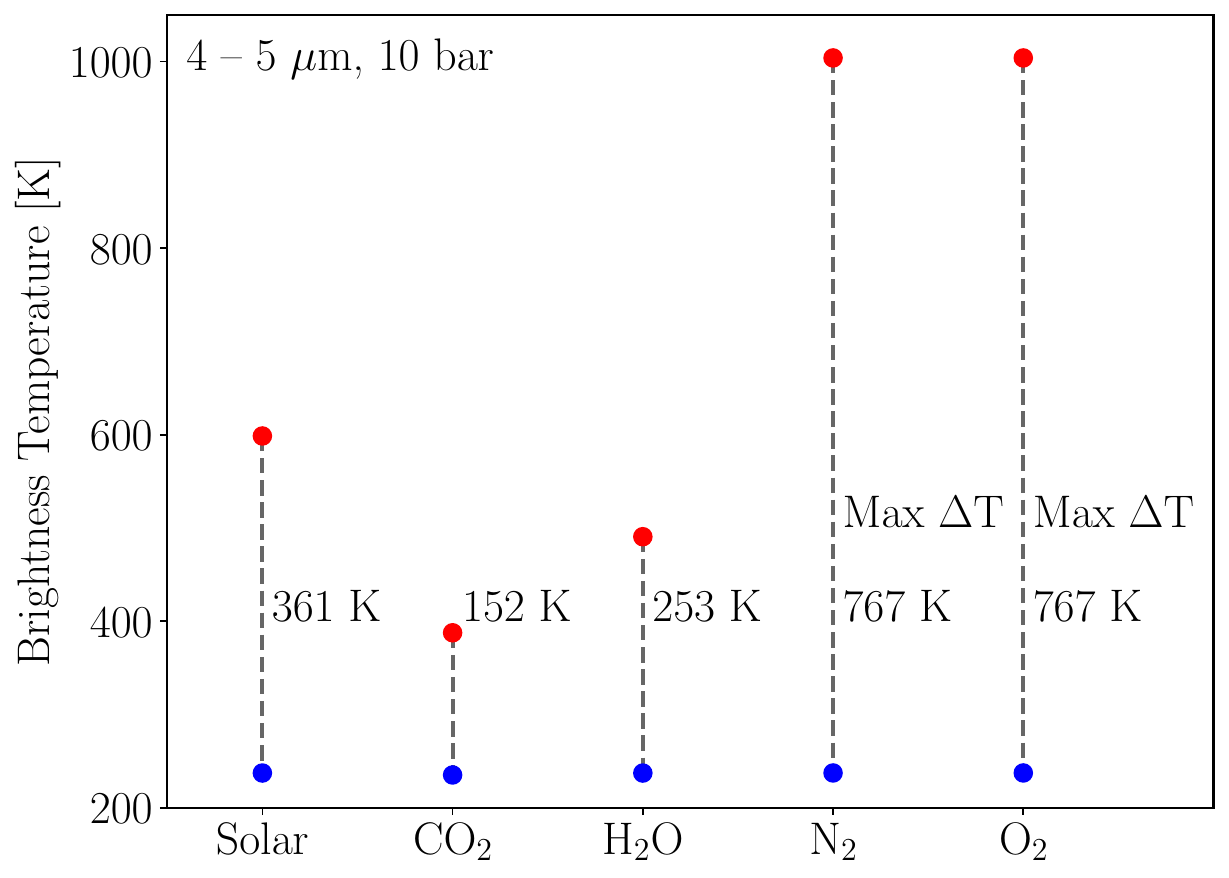}
    \includegraphics[width=0.49\textwidth]{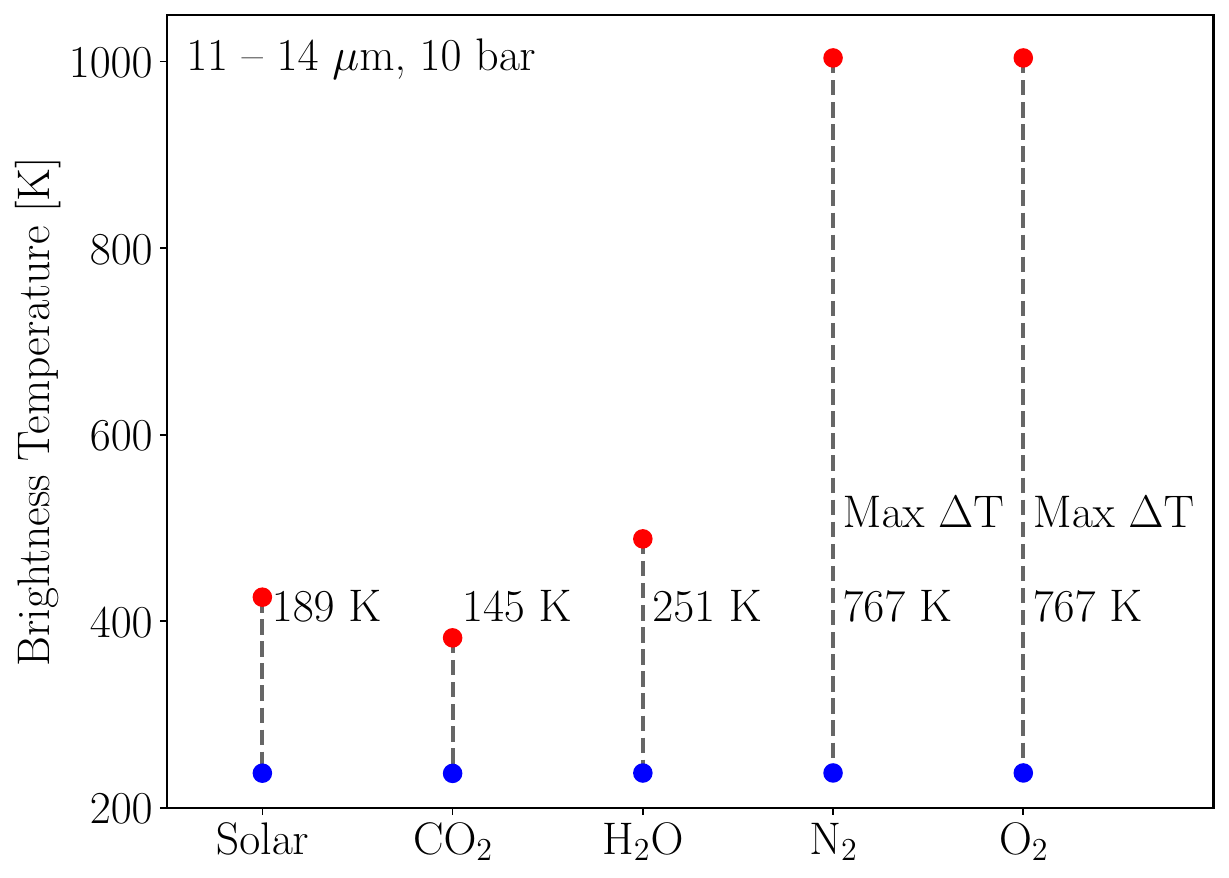}
   \includegraphics[width=0.49\textwidth]{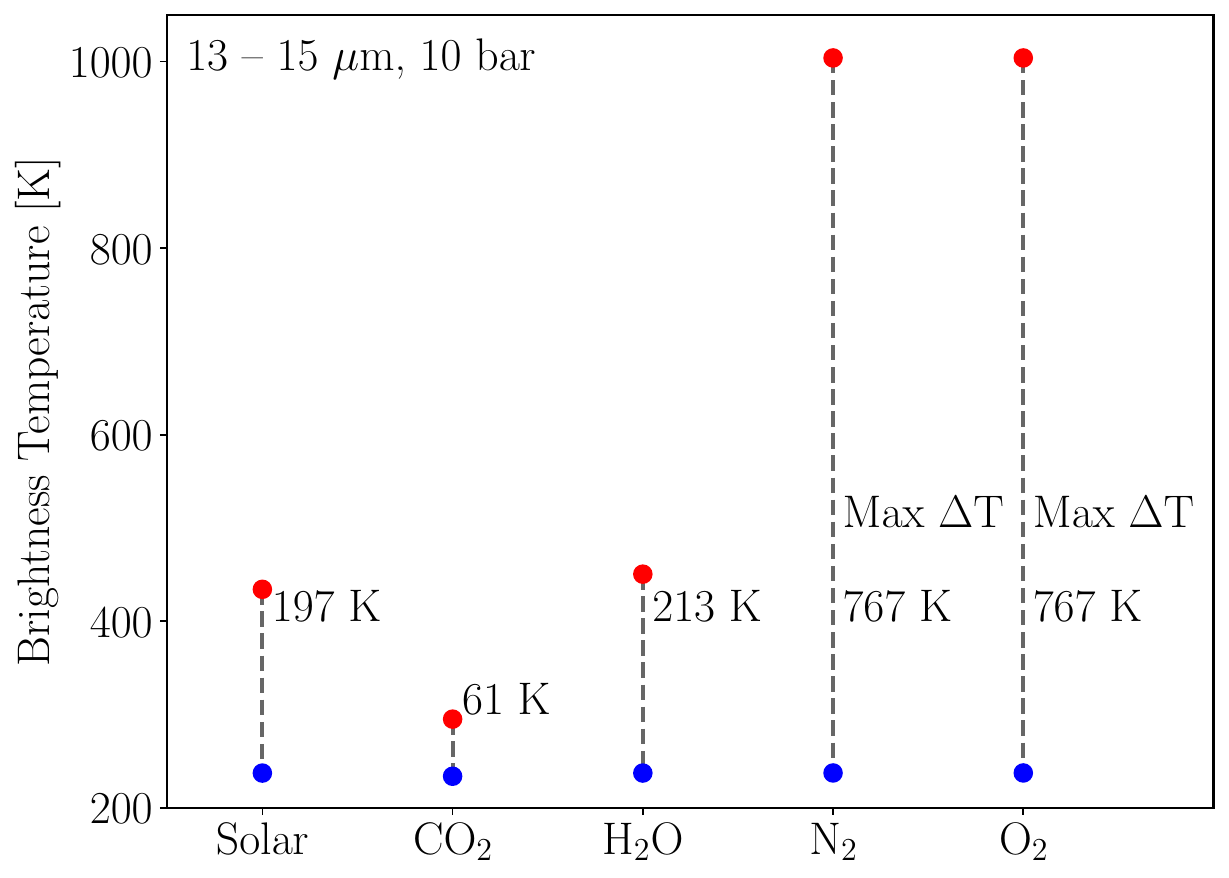}
   \caption{The dayside/nightside temperature contrast as a function of atmospheric composition as seen in observations averaged over 3-4 microns (top left)\DIFdelbeginFL \DIFdelFL{and }\DIFdelendFL , 4-5 microns (top right), 11-14 microns (bottom left), and 13-15 microns (bottom right). All atmospheres show dayside/nightside temperature contrasts that are enhanced compared to the true difference in dayside/nightside temperature (i.e., 50 K). The N$_2$ and O$_2$ dominated atmospheres show the most extreme observed dayside/nightside temperature contrast -- equivalent to the difference in temperature from the top to the bottom of the atmosphere. }
   \label{fig:diff_10bar}
\end{figure*}

\begin{figure*} 
   \centering
   \includegraphics[width=0.49\textwidth]{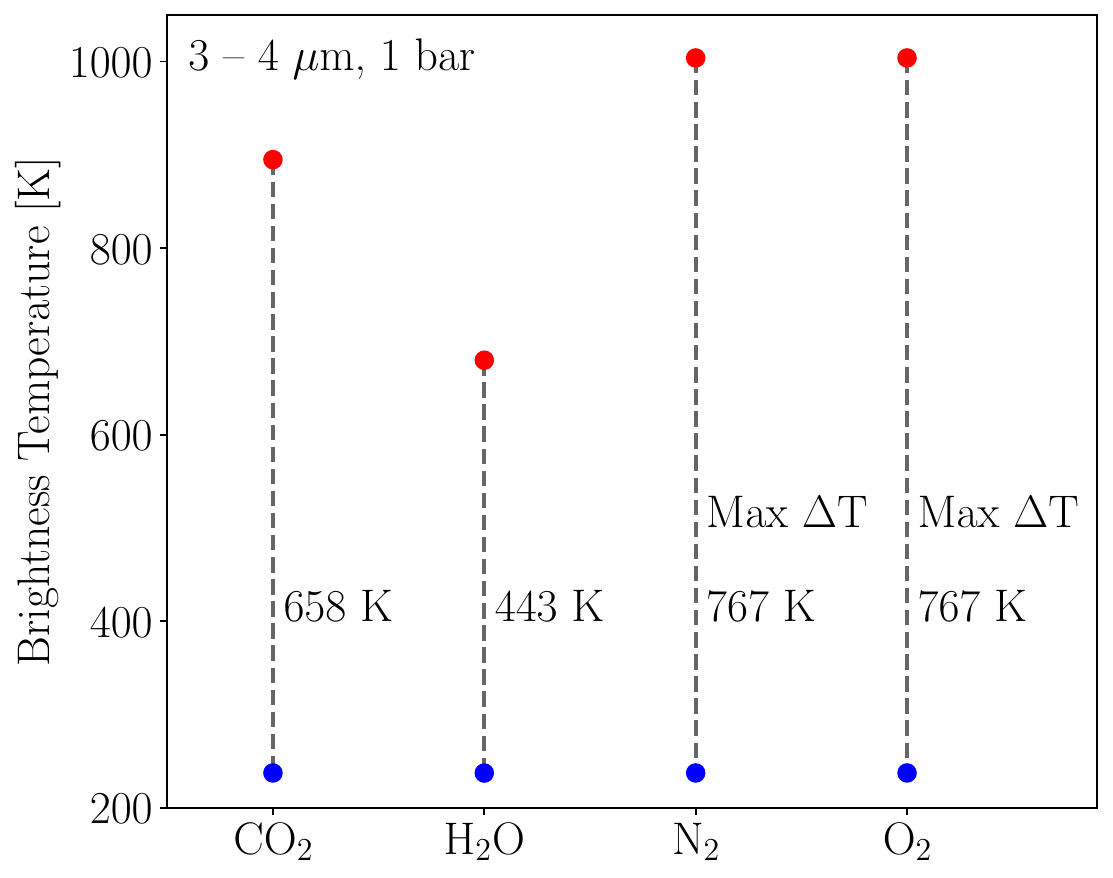}
   \includegraphics[width=0.49\textwidth]{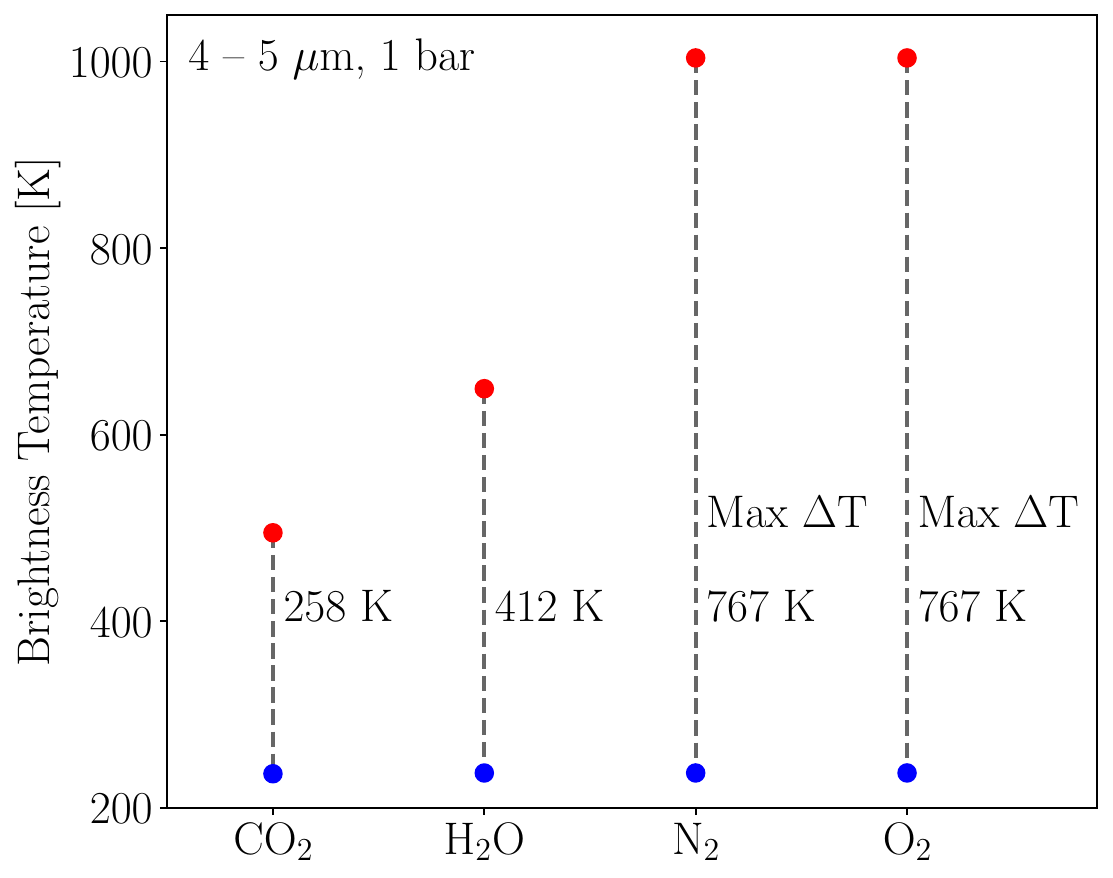}
   \includegraphics[width=0.49\textwidth]{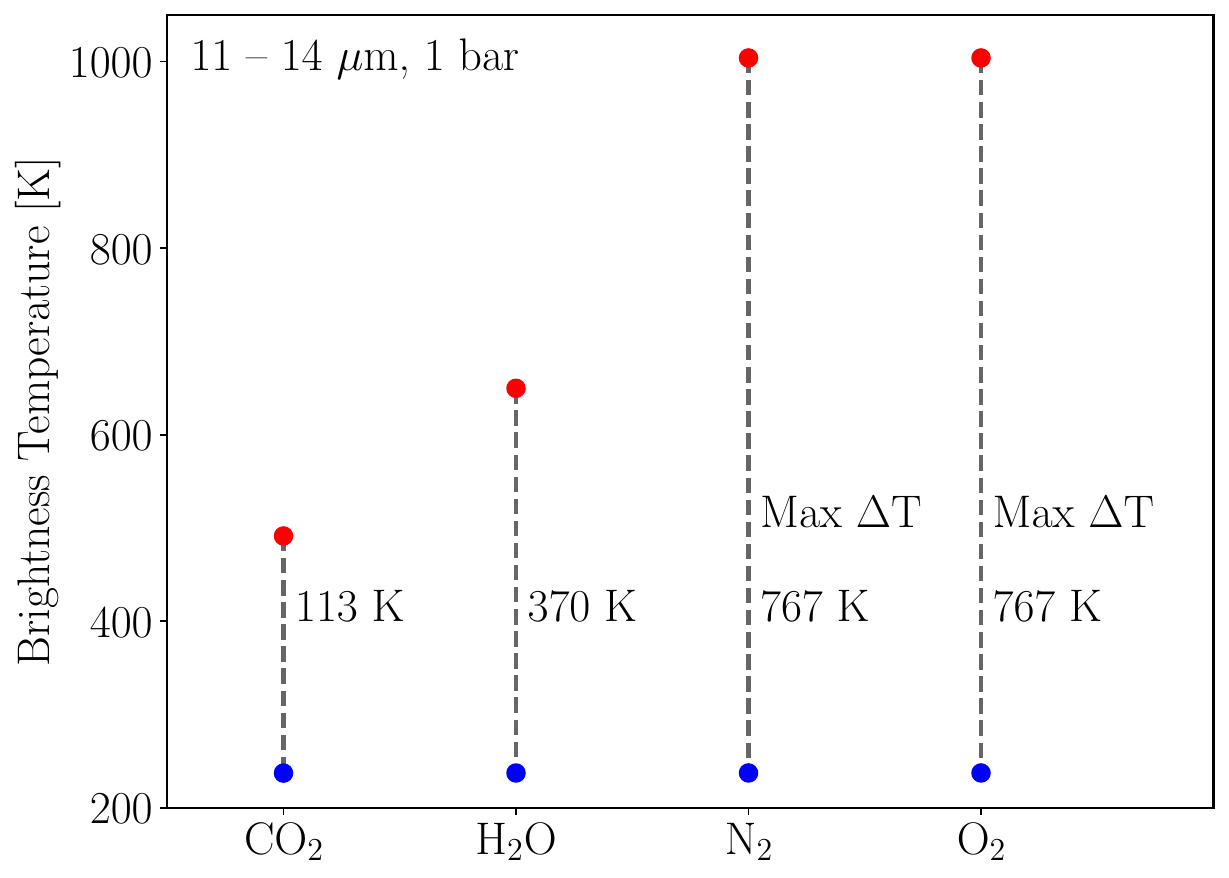}
   \includegraphics[width=0.49\textwidth]{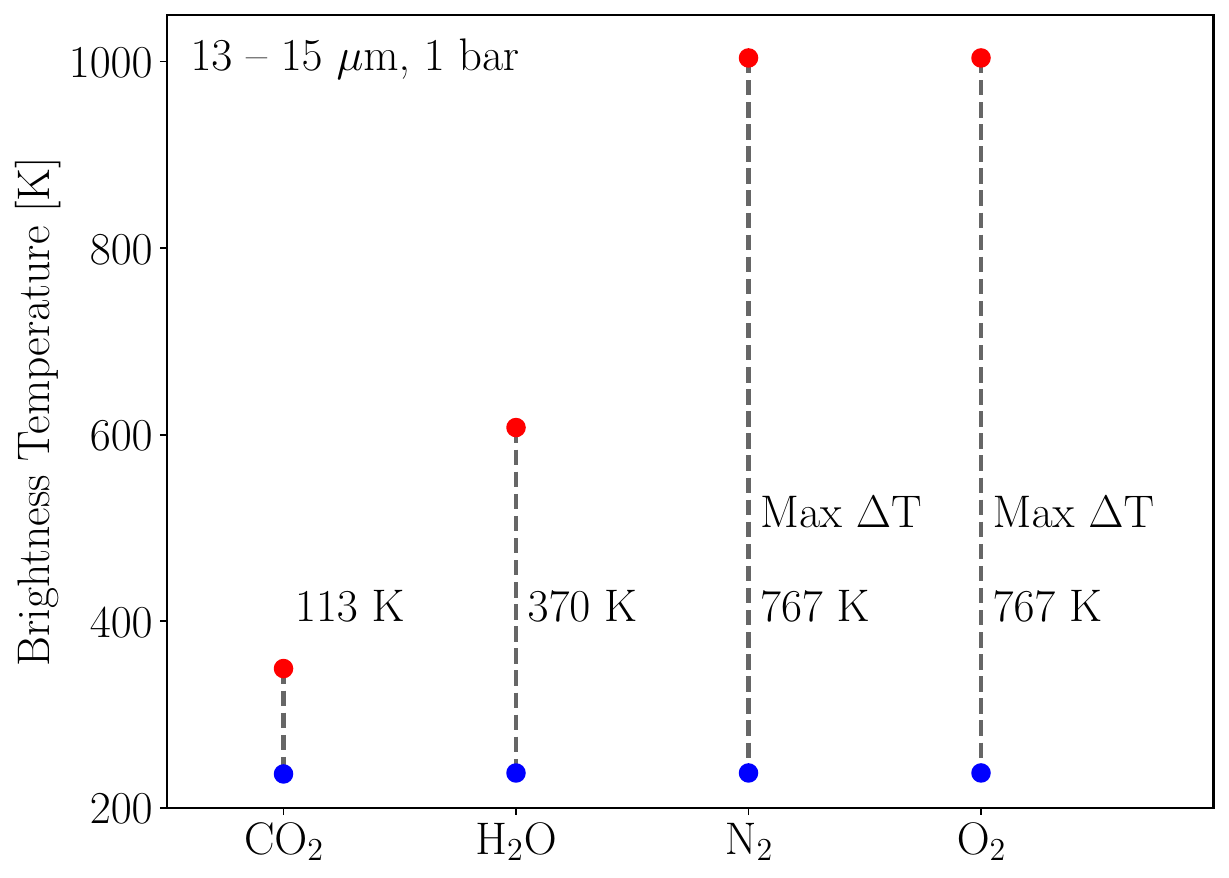}
   \caption{Same as Fig \ref{fig:diff_10bar} but for the 1 bar atmosphere case. In the case of a 1 bar planetary atmosphere, the dayside/nightside temperature contrast is exaggerated as compared to the 10 bar case. }
   \label{fig:diff_1bar}
\end{figure*}

We note that, in the case of atmospheric thermal structures similar to our fiducial case, nightside-only clouds are likely to occur in the case where there is a marginally supersaturated region in the upper atmosphere of the planetary nightside such that the \DIFdelbegin \DIFdel{condensible }\DIFdelend condensable species is undersaturated on the planetary dayside. Clouds that form in the deeper atmosphere are likely to be comprised of a \DIFdelbegin \DIFdel{condensible }\DIFdelend condensable species that condense on both the day and nightsides and, depending on the intricacies of the atmospheric dynamics, may homogenize the day/night brightness temperatures instead of exaggerate them. 


\section{The Impact of Nightside Clouds on Observed Day-Night Temperature Contrasts}\label{clouds_daynight}
We now more quantitatively investigate the impact of nightside clouds on the observed day-night temperature contrast of tidally locked terrestrial planets. We will consider a case of a relatively hot terrestrial planet. We again consider the fiducial 10 bar atmosphere case shown in Figure \ref{fig:temp}. We consider five different simplified atmospheric compositions which are commonly considered in the literature: solar, CO$_2$ dominated, H$_2$O dominated, pure N$_2$, and pure O$_2$. For simplicity, we consider pure atmospheric compositions though we note that some level of contribution in compositon from a  condensable species is required for more self-consistent modeling.

For our fiducial model we assume that an optically-thick gray cloud layer forms at $\sim$10$^{-2}$ bar on the nightside and extends over one dex in pressure. This cloud layer nominally has the condensation temperature of water, however, for the purposes of this study the exact cloud composition is unimportant. We discuss potential cloud compositions on terrestrial planets in more length in Section \ref{diverse_clouds}. The \DIFdelbegin \DIFdel{condensible }\DIFdelend condensable species is undersaturated on the planetary dayside which is thus cloud-free. 

In addition to our nominal case, we also consider the case of a 1 bar atmosphere with clouds present at the same atmospheric temperature and supersaturation (now at $\sim$10$^{-3}$ bar) with the same thermal structure except that the maximum temperature is present at 1 bar and we correspondingly model the atmosphere out to lower pressures. 

We use the \textit{PICASO} radiative transfer code \citep{Batalha2019} to model the dayside and nightside fluxes at 3-15 microns. We choose this range of wavelengths because this is the range used to determine whether a terrestrial planet hosts an atmosphere as planets are likely to emit the bulk of their emission at these wavelengths and they are accessible with JWST using the MIRI LRS observing mode. We use the system parameters assuming that our model planet orbits a star like LHS 3844 and has planet properties similar to LHS 3844b(Planet Mass = 0.003146 M$_\mathrm{Jup}$, Planet Radius = 0.08921 R$_\mathrm{Jup}$, T$_\mathrm{eff}$ = 3500 K, Stellar Radius = 0.178 R$_\odot$). The references for the molecular opacities are in Table~\ref{tab:opas} and we assume a surface with zero reflectivity.

We find that the dayside flux at 3-5 $\mu$m is many orders of magnitude larger than the nightside flux for all of the model atmospheres considered as shown in Figure \ref{fig:flux}. In the case of a H$_2$O, CO$_2$, or solar composition atmosphere, the dayside spectra exhibits significant atmospheric features at relevant wavelengths. In the case of the N$_2$ and O$_2$ dominated atmospheres the dayside spectra resembles a blackbody as there are no significant atmospheric features at these wavelengths given that our simplified atmospheric models neglects the diversity of unknown trace species that may be present. The nightside of the atmospheres in all cases resembles a blackbody with much less flux emitted than the planetary dayside. 

This flux difference translates into significantly exaggerated observed dayside/nightside brightness temperatures for every case we test. The differences in dayside/nightside brightness temperatures are shown in Figures \ref{fig:diff_10bar}-\ref{fig:diff_1bar}. Though the actual difference in dayside/nightside temperatures in our model atmospheres is only 50K, the range in observed dayside/nightside brightness temperatures is anywhere from 152 K to 767 K, where the high end of temperature difference represents the difference between the highest atmospheric temperature (near the surface) and the lowest atmospheric temperature (in the upper atmosphere). For both the 1 bar and 10 bar atmospheres, the N$_2$ and O$_2$ dominated atmospheres represent the largest difference in observed dayside/nightside brightness temperatures. This is because both the N$_2$ and O$_2$ dominated atmospheres do not absorb significantly at the infrared wavelengths of interest. Thus, on the planetary dayside, observations will probe emission from the planetary surface. In our model atmospheres the surface is $\sim$1000 K and the brightness temperature is a correspondingly similar value. As is the case for all model atmospheres probed, the planetary nightside brightness temperature originates from the top of cloud deck and emits at a bit more than $\sim$ 200 K. This cool nightside brightness temperature is undetectable with JWST. Thus, observations of a substantial (though optically-thin) N$_2$ or O$_2$ dominated atmosphere would appear as if the planet does not have an atmosphere at all.


\section{Potential Diversity of Clouds in Terrestrial Planet Atmospheres}\label{diverse_clouds}

Thus far we have spent little time on the particular cloud species that may form on the planetary nightside and give rise to such an observable signature. Rather we have adopted the condensation temperature corresponding to water clouds and assumed that the clouds are gray absorbers without any consideration to their formation mechanisms or microphysical properties which are beyond the scope of this work. However, we note that terrestrial planets are likely able to form a large diversity of clouds and cloud species, including novel species that may arise due to peculiar atmospheric compositions as is the case for hot Jupiters where cloud species such as SiO$_2$ are able to form \citep{grant2023}. 

\begin{figure} 
   \centering
   \includegraphics[width=0.49\textwidth]{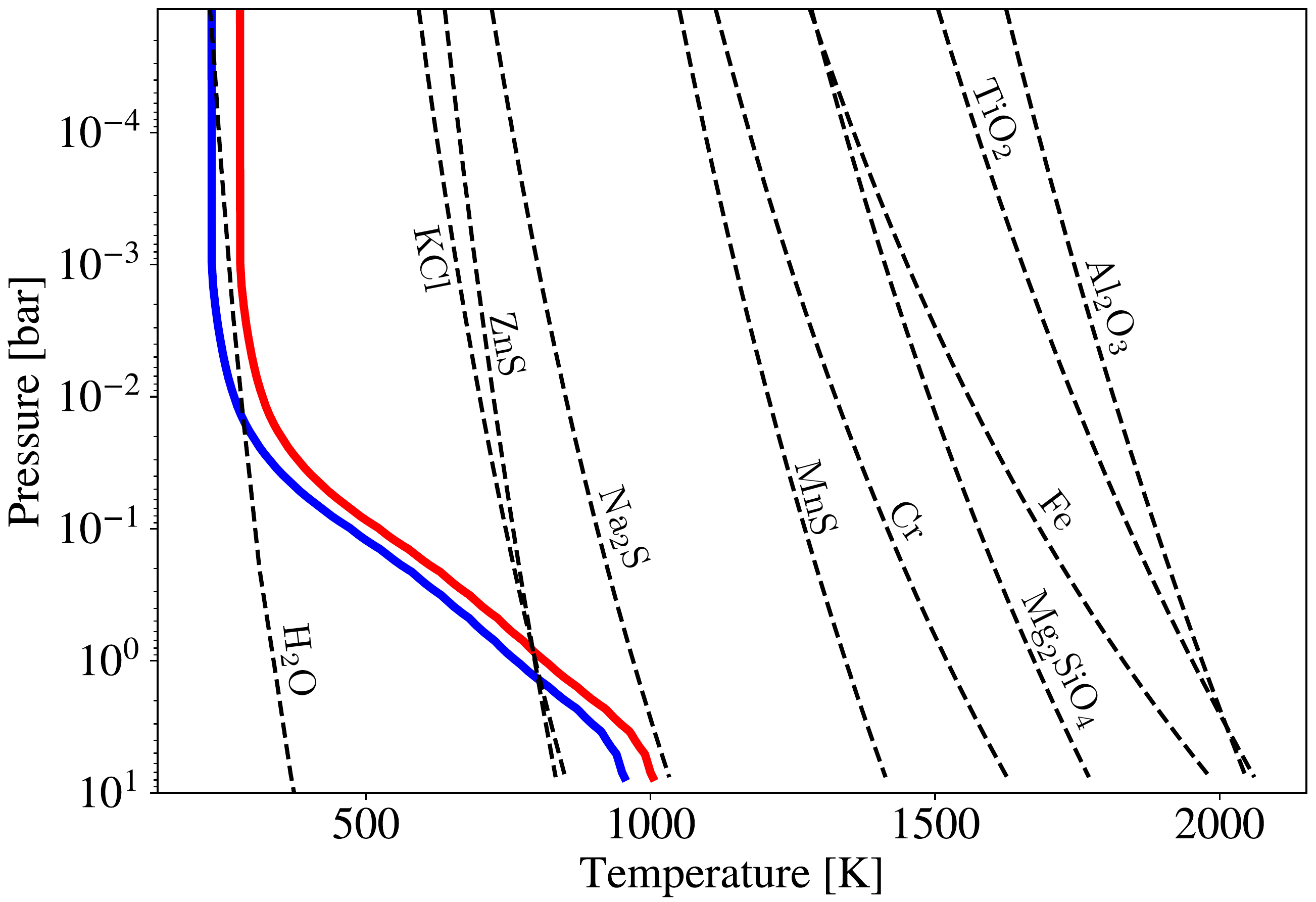}
   \caption{The condensation sequence for a solar composition atmosphere where dashed lines indicate the condensation curves for a variety of typically assumed \DIFdelbeginFL \DIFdelFL{condensible }\DIFdelendFL condensable species. The model planet atmosphere day (red) and night (blue) temperature profiles are shown. In the case of a primordial solar composition atmosphere there may exist a favorable region of temperature phase space where we do not expect significant opacity from condensational clouds (at $\sim$ 400 - 800 K). However, given the presumed diversity of terrestrial atmospheres, such a favorable region may be unlikely to exist.}
   \label{fig:cond_curves}
\end{figure}

In the case where a terrestrial planet hosts a nearly solar composition (primordial) atmosphere then there are a variety of cloud species that may dominate as a function of equilibrium temperature (see Fig \ref{fig:cond_curves}). This case would be analogous to giant exoplanet atmospheres such as hot Jupiters. The standard condensation sequence for solar metallicity atmospheres described in Figure \ref{fig:cond_curves} primarily consists of high temperature metal clouds, followed by intermediate temperature sulfur-bearing condensates and salts, and then the cooler condensation of water clouds. Many of these cloud species aside from silicates and water clouds may have little impact on atmospheric spectra \citep[e.g.,][]{2020NatAs...4..951G,Gao2021}. Thus, when considering these cloud species previously speculated in the literature, there may be no spectrally dominant cloud species that form at roughly 400-1500 K such that this region of phase space may be favorable for observing cloud-free atmospheres. 

However, the atmospheric compositions of evolved terrestrial planets are likely to vary substantially across the population and be different in composition to a solar-type composition atmosphere. More work that addresses a diversity of atmospheric compositions and potential aerosols given those compositions is thus essential in better understanding terrestrial exoplanets \citep[along the lines of][]{Mbarek2016}. 

\section{Discussion}\label{discuss}

\begin{figure} 
   \centering
   \includegraphics[width=0.49\textwidth]{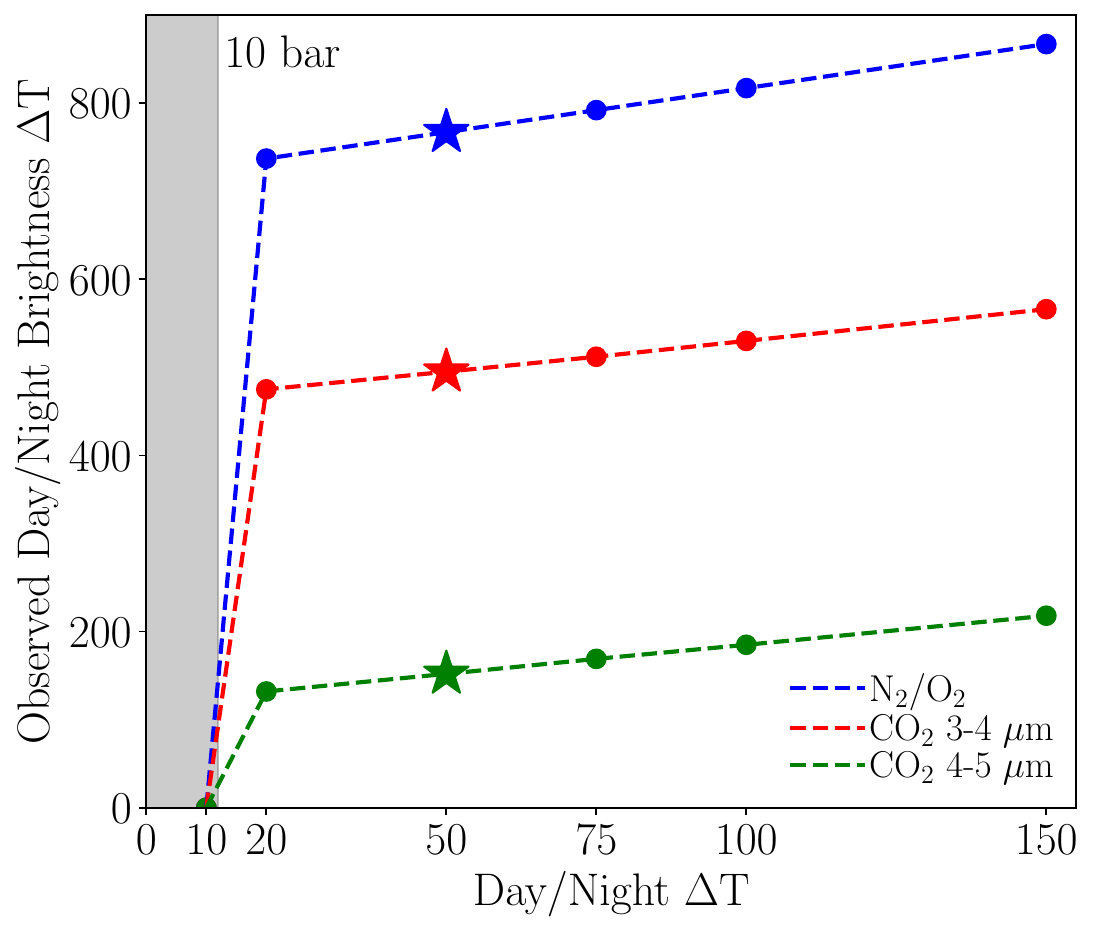}
   \caption{The observed contrast in dayside/nightside brightness temperature as a function of actual day/night temperature contrast for a 10 bar atmosphere with an N$_2$ or O$_2$ dominated atmosphere (observed at either 3-4 or 4-5 $\
   mu$m) and a CO$_2$ dominated atmosphere observed at two different infrared wavelength ranges. For our fiducial temperature profile structure, the observed day/night brightness temperature contrast remains roughly constant for actual atmospheric temperature contrasts greater than 10K. For temperature contrasts less than 10K (gray region), the observed brightness temperature contrast depends on the efficiency of cloud formation on the dayside given a very marginal \DIFdelbeginFL \DIFdelFL{condensible }\DIFdelendFL condensable supersaturation and thus requires more detailed microphysical modeling to reach robust conclusions. }
   \label{fig:int_daynight}
\end{figure}

This work has sought to demonstrate additional complexity in terrestrial atmospheres that needs to be considered when interpreting observations of terrestrial planets in emission. In particular, we aim to demonstrate that there is a vast and relatively unexplored region of phase space relevant to interpreting current observations. In this effort we have considered a simplified picture which requires more detailed follow-up study.

In particular, we have made several simplifying assumptions regarding the atmospheric temperature/pressure profile and the level of intrinsic day/nightside temperature contrast. In Figure \ref{fig:int_daynight}, we demonstrate how the observed day/nightside brightness temperature difference depends on the intrinsic day/nightside temperature difference. We find that for actual day/nightside temperature contrasts that are greater than 10 K, the observed difference in day/night brightness temperatures remains constant, thus this picture is relatively insensitive to the magnitude of the day/night temperature contrast as long as the dayside is too hot to form the cloud in question. For day/night atmospheric temperature differences of less than 10 K, the observed brightness temperature contrast depends on the efficiency of cloud formation on the atmospheric dayside which may be marginally supersaturated. In this case, more detailed dynamical, radiative, and microphysical modeling is needed to reach robust conclusions. More generally, the probability that a cloud species' condensation temperature is \DIFdelbegin \DIFdel{in between }\DIFdelend in-between the day and nightside atmospheric temperature decreases with decreasing intrinsic temperature contrast. Thus, the likelihood of this effect depends on the intrinsic magnitude of the dayside/nightside atmospheric temperature difference.   

There may be several additional complications that could play a first order role in altering this picture, including: the shape of the atmospheric temperature/pressure profiles, \DIFdelbegin \DIFdel{the atmospheric radiative transfer, }\DIFdelend interplays between dynamics, cloud formation, and radiative transfer, and dayside photochemistry/haze production. For example, a local dayside thermal inversion due to the presence of a photochemically produced species may increase the day/night temperature contrast and lead to an exacerbated observed brightness temperature contrast (i.e., both a hotter dayside and cooler nightside). Additionally, near-IR absorption of stellar radiation alone may lead to non-adiabatic temperature profiles that could inhibit cloud formation on rocky planet daysides in general, particularly for M-dwarfs \citep{malik2019}. Conversely, a nightside atmosphere that is hotter above the optically thick cloud deck would increase the observed nightside brightness temperature and could decrease the day/night brightness temperature contrast. There is also the potential for dayside hazes to form on warm/hot terrestrial planets. In the case of significant haze formation on the planetary dayside, the planet is likely to resemble the case of an efficient dayside cold trap and appear cooler than its intrinsic temperature \DIFdelbegin \DIFdel{and }\DIFdelend such that the planet would be harder to detect in emission. \DIFdelbegin \DIFdel{And furthermore}\DIFdelend Furthermore, the interplay between cloud formation, atmospheric dynamics, and radiative transfer, may lead to spatially inhomogeneous cloud formation \citep[as in][]{Turbet2021} across a significantly broader range of terrestrial atmospheres than those considered in this work. 

Many observing programs with \textit{JWST} seek to establish the existence or lack of an atmosphere by observing the dayside brightness temperature alone which has been theorized to be a robust probe of the existence of an atmosphere. However, if the formation of nightside clouds has a similar radiative impact on terrestrial atmospheres as it does on hot Jupiter \DIFdelbegin \DIFdel{atmospheres where }\DIFdelend atmospheres---where it increases the dayside atmospheric temperature \citep{parmentier2021}\DIFdelbegin \DIFdel{, then }\DIFdelend ---then the interpretation of dayside brightness temperatures needs to include a consideration of cloud radiative effects. 

Taken together, future work that self-consistently models the atmospheric temperature structure, dayside/nightside temperature contrast, cloud/haze microphysics, the interplay between cloud formation and atmospheric dynamics, and the radiative influence of nightside clouds on the atmospheric temperature structure are needed to better understand the potential impact of clouds on atmospheric processes and observations.

\section{Conclusions}\label{conclude}

We have demonstrated that nightside clouds can give the illusion of extreme day/night temperature contrasts when in reality the temperature contrast may only be 10s of K. We thus note that a significant dayside/nightside temperature difference does not by itself indicate that the planet is atmosphere-free and instead indicates the need for more detailed models and spectroscopic observations of the planetary atmosphere. We note, however, that a planet with a homogeneous atmospheric temperature structure almost certainly has an atmosphere as discussed extensively in the existing literature. In both cases, atmospheric spectra are needed to make robust conclusions about planetary characterization. 

More detailed models and GCM studies with varying volatile inventories and equilibrium temperatures will help to better understand the potential properties of hot/warm terrestrial planet atmospheres. In particular, future work that investigates the diversity of aerosol species that may form in hot/warm terrestrial atmospheres, the radiative effects of clouds, the interplay between clouds and atmospheric dynamics, and the formation and evolution of clouds and hazes will be important to understand observations of these complex worlds.

\appendix
The molecular opacities used to generate the emission spectra in this work are taken from the references in Table \ref{tab:opas}. 

\begin{deluxetable*}{lc}[!htbp]
\tablecolumns{2}
\tablehead{   
  \colhead{Species} &
  \colhead{Reference} 
}
        \centering
        \startdata
        CO2 &  \citet{HUANG2014reliable} \\ 
        CH4 &  \citet{yurchenko13vibrational,yurchenko_2014} \\ 
        CO &  \citet{HITEMP2010,HITRAN2016,li15rovibrational} \\ 
        K &  \citet{Ryabchikova2015,Allard2007AA,Allard2007EPJD,Allard2016,Allard2019} \\ 
        H2 &  \citet{HITRAN2016} \\ 
        H2O &  \citet{Polyansky2018H2O} \\ 
        H2S &  \citet{azzam16exomol} \\ 
        H2--H2 &  \citet{Saumon12,Lenzuni1991h2h2} \\ 
        H2--He &  \citet{Saumon12} \\ 
        H2--H &  \citet{Saumon12} \\ 
        H2--CH4 &  \citet{Saumon12} \\ 
\enddata
            \caption{Line lists used to make PICASO Opacities}
            \label{tab:opas}
        \end{deluxetable*}

\bibliography{references}{}
\bibliographystyle{aasjournal}

\end{document}